\documentclass[11pt]{article}
\usepackage{moriond,epsfig}

\def\Journal#1#2#3#4{{#1} {\bf #2}, #3 (#4)}

\def\PRD{{\em Phys. Rev.} D}

\def\be{\begin{equation}}
\def\ee{\end{equation}}
\def\bea{\begin{eqnarray}}
\def\eea{\end{eqnarray}}

\def\tt{$t\bar{t}$}
\def\pp{$p\bar{p}$}
\def\pb{pb$^{-1}$}

\begin{document}


\vspace*{4cm}
\title{TOP PHYSICS RESULTS FROM CDF}

\author{ GERVASIO G\'OMEZ }

\address{for the CDF Collaboration \\
Instituto de F\'{\i}sica de Cantabria (CSIC-Universidad de Cantabria) \\
Av. de los Castros s/n \\
39005 Santander, Spain}

\maketitle\abstracts{
The top quark is by far the most massive fundamental particle observed so far, 
and the study of its properties is interesting for several reasons ranging
from its possible special role in electroweak symmetry breaking to its 
sensitivity to physics beyond the Standard Model.
We present recent top physics results from CDF based on 160-320 \pb\
of \pp\ collision data at $\sqrt{s}=1.96$ TeV. 
The \tt\ cross section and the top mass have been measured in different 
decay channels and using different methods. We have searched for evidence of 
single top production, setting upper limits on its production rate. Other 
results shown in this conference include studies of the polarization of $W$ 
bosons from top decays, a search for charged Higgs decaying from top, and a 
search for additional heavy $t^{\prime}$ quarks.}

\section{Introduction: top at CDF}
The CDF detector was upgraded~\cite{cdf} for Run 2 of the Tevatron, and has
recorded $\approx$ 600 \pb\ of \pp\ collision data at 
$\sqrt{s}=1.96$ TeV. Top at the Tevatron is produced predominantly in 
\tt\ pairs through the strong interaction (quark-antiquark annihilation 
or gluon gluon fusion), and it decays virtually 100\% of time to $Wb$.
The final state of a \tt\ event therefore has two $W$s and two 
$b$ quarks, and the event selection at CDF is characterized by the decay mode
of the $W$ bosons. Events where both $W$s decay to $e$ or $\mu$ are called
"dilepton" events. This mode has the advantage of being relatively clean, 
with a S/N of about 1.5 to 3.5, but it has a low rate 
(4-6 events/100 \pb) owing to the small leptonic branching fraction of 
$W$. Events in which one $W$ decays to $e$ or $\mu$ and the other decays to
quarks are called "lepton + jets" events. This decay channel has a higher rate 
than the dilepton channel (25-45 events/100 \pb), but it has worse S/N
(0.3 to 3). Since all \tt\ events have two $b$ quarks while only 1-2\% of
the dominant backgrounds contain heavy flavor, S/N can be considerably improved
by identifying $b$ quarks ($b$ tagging). This is done either by looking at 
tracks embedded in a jet which are displaced from the primary interaction 
point (due to the long $b$-hadron lifetime) or by looking for a soft lepton 
embedded in a jet (due to the large semileptonic decay rate of $b$-hadrons).
Top at the Tevatron can also be produced singly through the electroweak 
interaction, with a predicted cross section which is about 3 times smaller than
the \tt\ cross section. Single top events contain just one $W$ in the 
final state, have lower jet multiplicity and are harder to separate from the
background. 

\section{Top Cross-Section}
Measurement of the \tt\ production cross section in the different decay modes 
and with different methods provides an understanding of the event selection 
efficiency, background contamination, event kinematics and heavy flavor 
content. It is the benchmark measurement to validate a sample of top events 
which can be used for other top quark measurements, and it is in itself a test 
of QCD predictions. 
The most accurate measurement in the dilepton channel results from the 
combination two complimentary analyses using approximately 200 \pb.
One analysis selects events with two identified leptons ($e$ or $\mu$), 
while the other identifies one lepton and requires an additional isolated high 
$P_T$ track in order to increase the acceptance at the cost of some additional 
background contamination. The two lepton analysis observes 13 events with a
background expectation of $2.7 \pm 0.7$, while the lepton+track analysis 
observes 19 \tt\ candidates with a background expectation of $6.9 \pm 1.7$.
The combined analysis gives
\bea
\sigma_{t\bar{t}} = 7.0^{+2.4}_{-2.1}\rm{(stat)}^{+1.6}_{-1.1}\rm{(sys)}
\pm 0.4\rm{(lum) \ pb.} \nonumber
\eea

\begin{figure}
\begin{center}
\psfig{figure=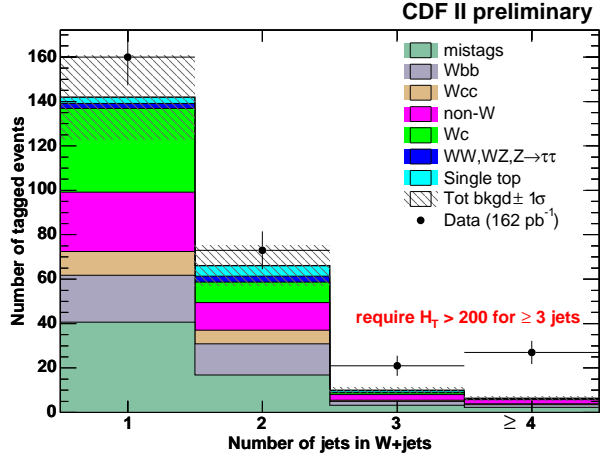,height=2.5in}
\caption{The number of $b$-tagged lepton+jets events and estimated background
as a function of jet multiplicity. 
\label{fig:btagxsec}}
\end{center}
\end{figure}

In the lepton+jets decay mode several cross section measurements have been 
performed. The basic event selection requires one identified lepton, large
missing transverse energy and 3 or energetic jets. Counting analyses
require at least one of the jets to be tagged as a $b$-jet in order to further
reduce the dominant $W$+jets and QCD backgrounds. The QCD and fake-tag 
backgrounds are derived from data, while $W$ + heavy flavor, single top, and
diboson backgrounds use a combination of data and MC. We check the background 
estimation in a control region of low jet multiplicity, where we expect little
\tt\ contribution, and measure the cross section the events with 3 or more 
jets by assigning any excess of observed events over the predicted non-top SM
backgrounds to \tt\ production. 
The most precise determination of the \tt\ cross section at CDF uses $\approx$
160 \pb\ of lepton+jets events and requires at least one jet to be $b$-tagged
by a secondary vertex algorithm which looks at tracks associated to the jet.
Additionally, the total transverse energy $H_T$ of the event 
(including the missing $E_T$) is required to be larger than 200 GeV in the 
signal region. 
Figure~\ref{fig:btagxsec} shows the distribution of $b$-tagged events and the
predicted background rates as a function of jet multiplicity. 
A total of 48 events are observed in the 3 or more jet region over an expected
background of $13.5 \pm 1.8$. The resulting cross section is
\bea
\sigma_{t\bar{t}} = 5.6^{+1.2}_{-1.1}\rm{(stat)}^{+0.9}_{-0.6}\rm{(sys)
 \ pb } \nonumber
\eea
where a 6\% luminosity uncertainty is included in the systematic uncertainty.

\begin{figure}
\begin{minipage}{0.45\linewidth}
\psfig{figure=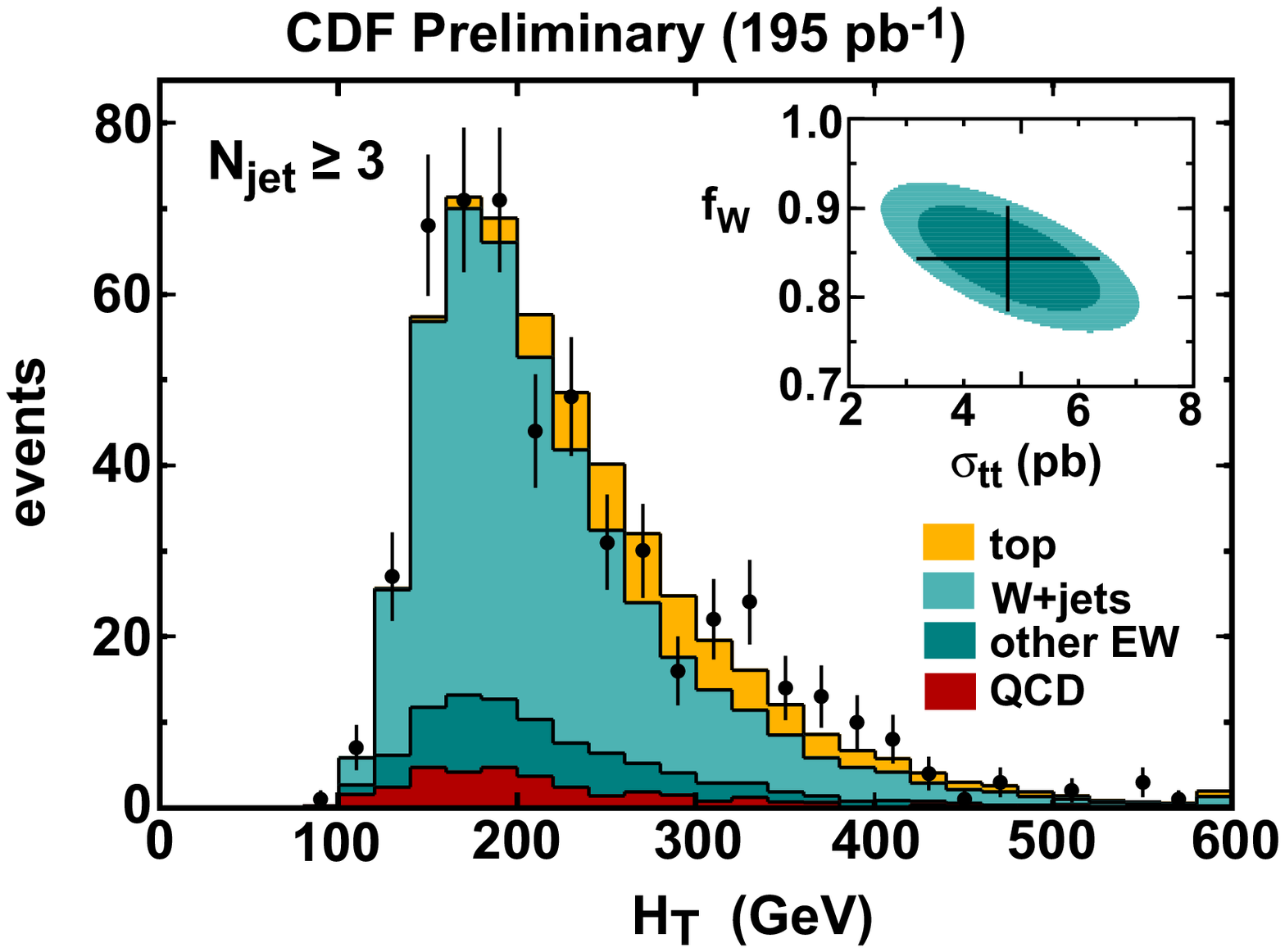,height=2.5in}
\end{minipage} \hfill
\begin{minipage}{0.45\linewidth}
\psfig{figure=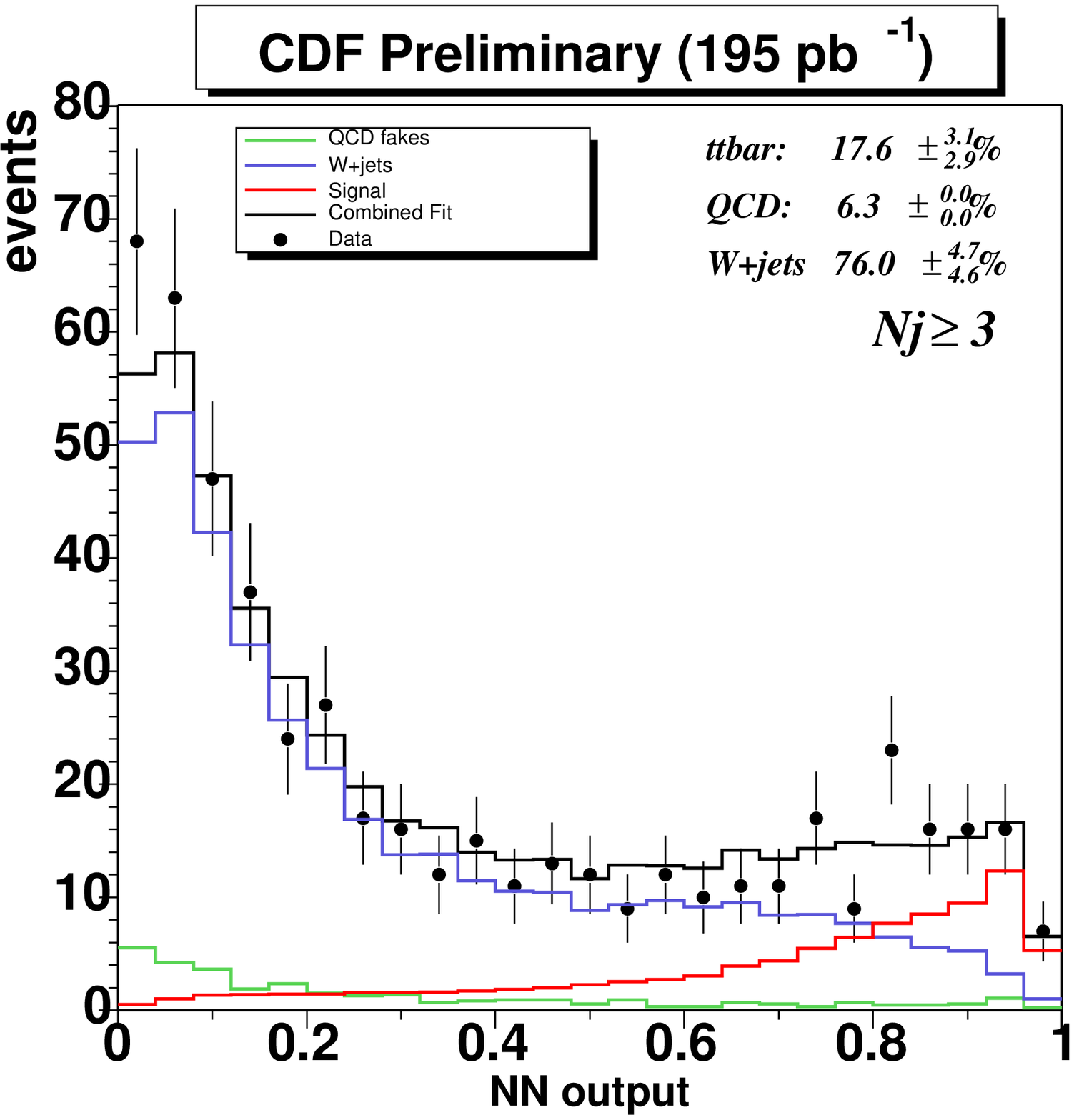,height=2.5in}
\end{minipage} \hfill
\caption{Distributions of $H_T$ (left) and NN output (right) for data, top and
background. Signal and background distributions are fit to the data to 
determine the \tt\ fraction. \label{fig:kinxsec}}
\end{figure}

Other analyses in the lepton+jets channel do not use $b$-tagging. Instead, 
they make use of kinematic information in the \tt\ candidate events. 
Templates for different kinematic variables are built for signal and background
events, and the distribution observed in the data is fit to a sum of these 
templates, allowing the normalizations to float. Two analyses of this type have
been performed at CDF. One simply fits the $H_T$ distribution, while the
other one derives a neural net (NN) discriminant which includes information 
from seven different kinematic distributions. The results of such fits are 
shown in Figure~\ref{fig:kinxsec}. The $H_T$ fit illustrates the difficulty of 
separating the top signal from the background when one does not use 
$b$-tagging. The fraction of top from this fit is ($13 \pm 4$)\%. The NN
analysis uses $\approx$ 195 \pb\ and observes 519 (pretag) events. The \tt\ 
fraction is ($17.6 \pm 3.0$)\% and the measured cross section is
\bea
\sigma_{t\bar{t}} = 6.7 \pm 1.1\rm{(stat)} \pm 1.6\rm{(sys)
 \ pb.} \nonumber
\eea

Several other measurements of the cross section have been performed which 
include two $b$-tagged jets, soft lepton tagging, all hadronic $W$ decays,
among others. All measurements are consistent with each other and with the
theoretical prediction~\cite{xsec-th}
$\sigma_{t\bar{t}}^{\rm{th}} = 6.7^{+0.7}_{-0.9}$ pb for $m_t = 175$ GeV.

\section{Top Mass}
One of the most interesting properties of top is its huge mass, so large that
its Yukawa coupling is stinkingly close to unity, making us wonder if
this is just a coincidence or if top plays some special role in the electroweak
symmetry breaking mechanism. The top mass is also a dominant parameter in 
higher order radiative corrections to several SM observables, and in particular
an accurate determination of the top mass, combined with precision electroweak
measurements, helps constrain the mass of the elusive SM Higgs.
CDF has measured the top mass using both dilepton and lepton+jets events.
These analyses are difficult for several reasons: the final state leptons
and jets observed in the detector must be matched to the \tt\ decay partons,
giving several possible assignments (only one of which is correct), the
undetected neutrinos cause the event kinematics to be under-constrained, and
the jet energies must be known with high accuracy. The dominant 
systematic uncertainty for all top mass measurements is the jet energy 
determination. As of the date of this conference, CDF is in the process of
revising the jet energy corrections, which will result in a reduction of the
systematic uncertainty of all $m_t$ results shown here to about one half of 
their value. Much improved $m_t$ results will be shown in the next round of
conferences. 

\begin{figure}
\begin{minipage}{0.43\linewidth}
\psfig{figure=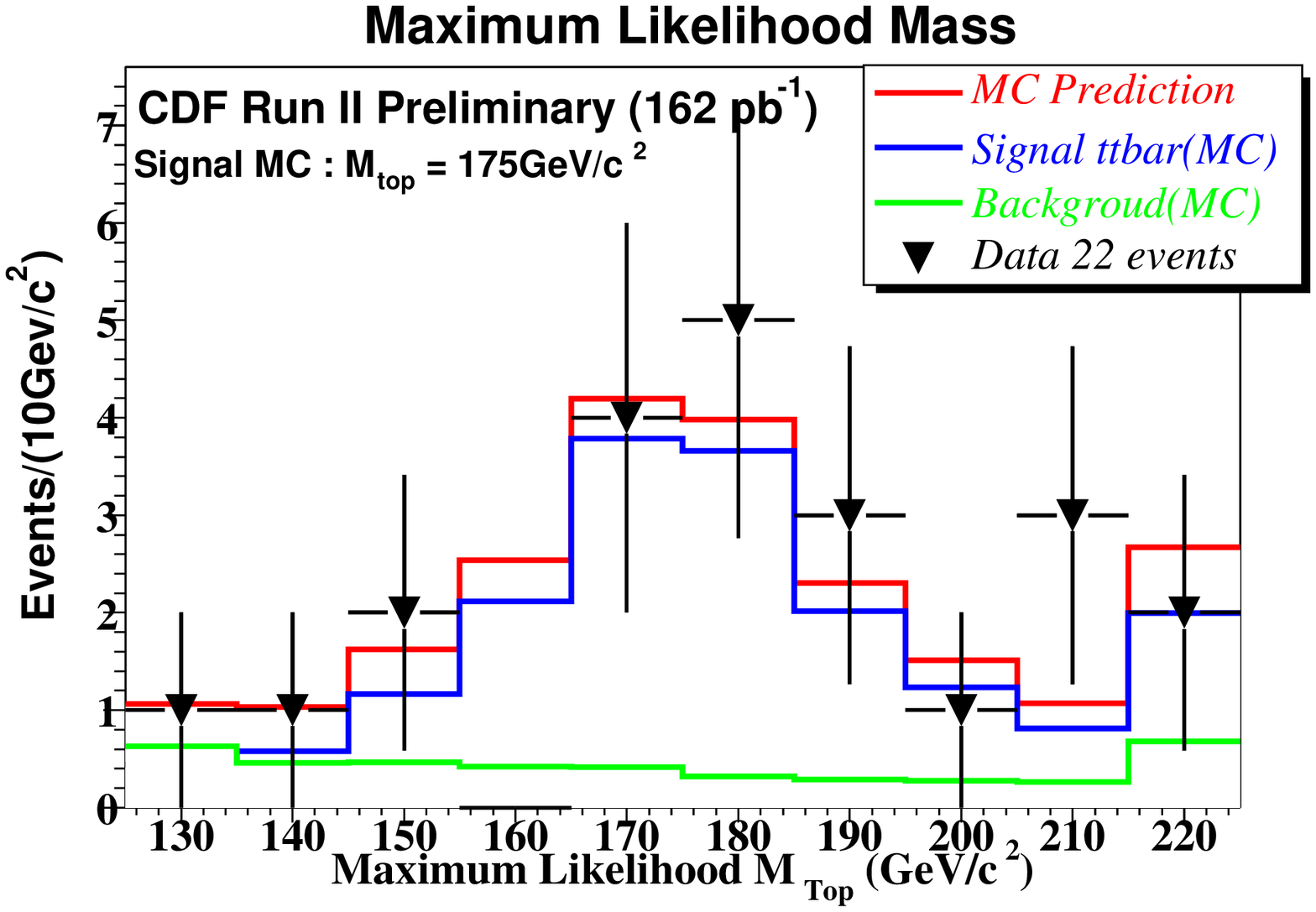, height=2.5in, width=\linewidth}
\end{minipage} \hfill
\begin{minipage}{0.43\linewidth}
\psfig{figure=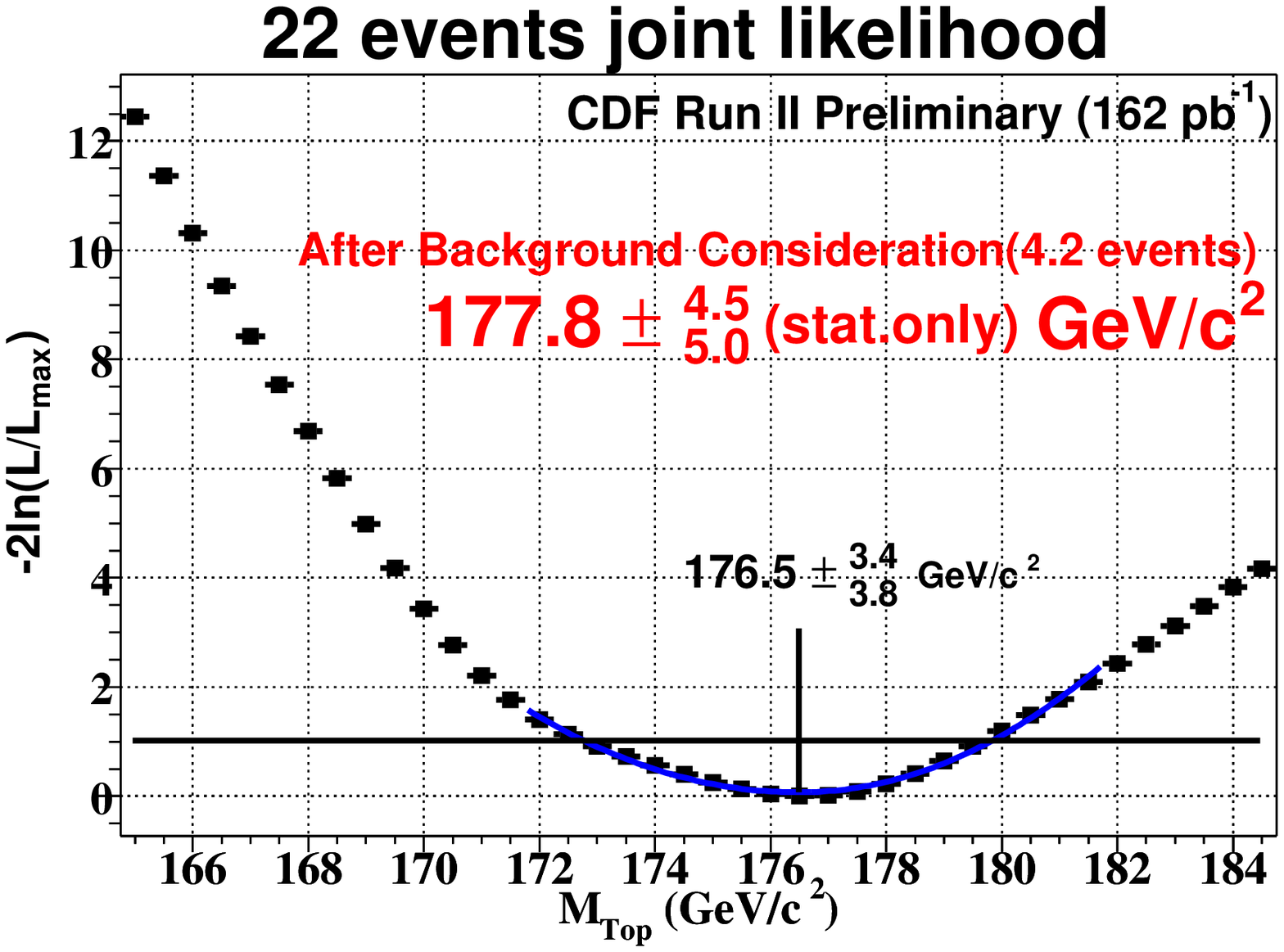,height=2.5in, width=\linewidth}
\end{minipage} \hfill
\caption{Maximum likelihood mass distribution for the 22 observed events (left)
and combined likelihood for all 22 events (right). \label{fig:dlm}}
\end{figure}

The best CDF result comes from a "dynamic likelihood method" (DLM) analysis 
which uses $\approx$ 160 \pb\ of tagged lepton+jets data. Event selection 
requires one identified $e$ or $\mu$ and exactly 4 jets, at least one of which 
is required to be tagged by the secondary vertex algorithm. A likelihood is 
built as a 
function of $m_t$ for each event using information from leading order matrix 
element (LO ME) Monte Carlo and transfer functions which give the probability 
of observing final state jets and leptons given the LO ME partons. All 
parton-jet assignments are considered, with the correct assignment receiving a 
larger weight. The top mass is obtained by maximizing the combined likelihood
of all 22 events passing the selection, and it is then corrected by a 
background mapping function which gives the true top mass for the measured
background fraction of 19\%. Figure~\ref{fig:dlm} shows the maximum likelihood
mass distribution for each of the 22 observed events and the resulting
combined likelihood. The top mass is determined to be
\bea
m_t = 177.8^{+4.5}_{-5.0}\rm{(stat)} \pm 6.2\rm{(sys) \ GeV. } \nonumber
\eea

\begin{figure}
\begin{minipage}{0.43\linewidth}
\psfig{figure=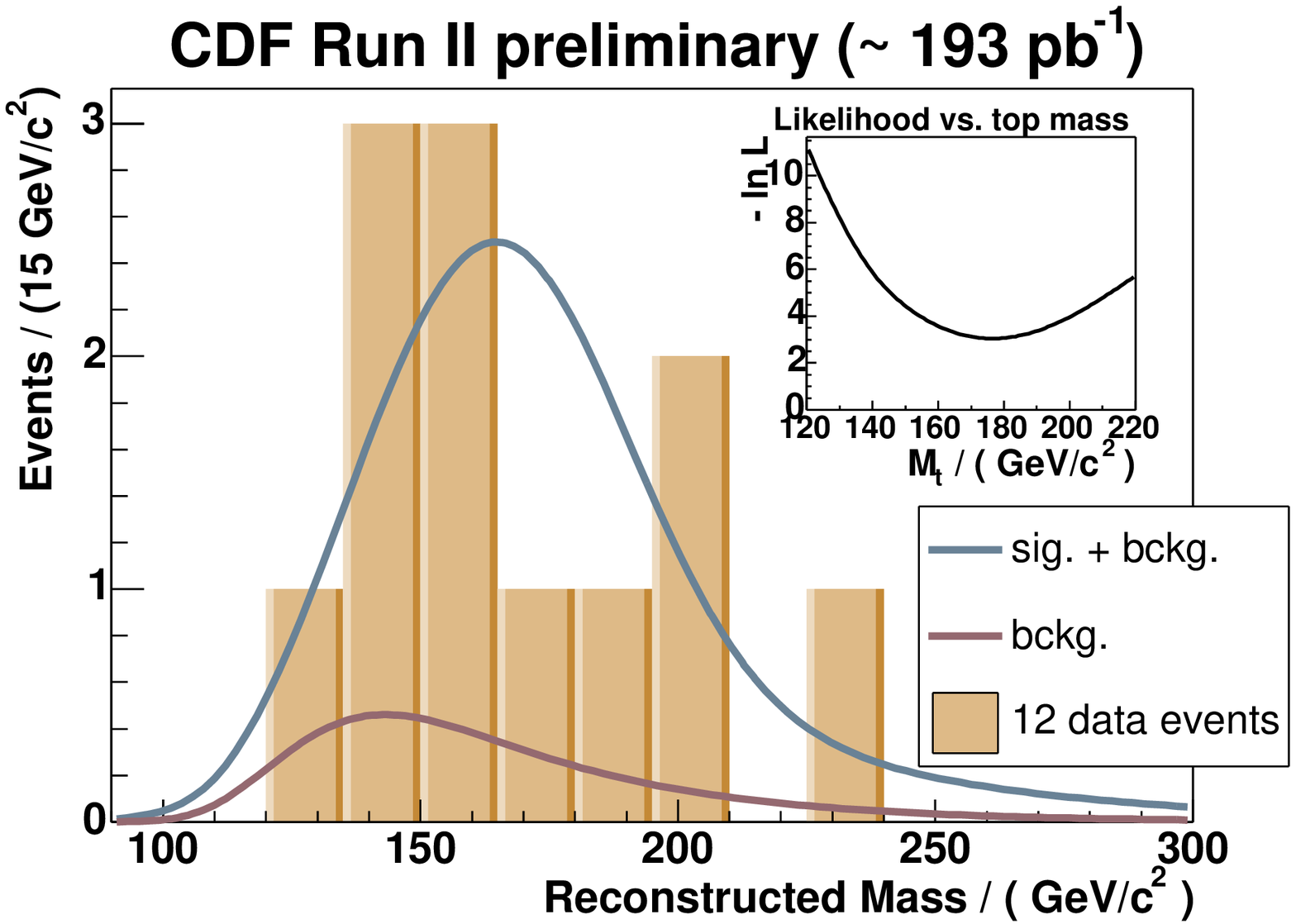, height=2.5in, width=\linewidth}
\end{minipage} \hfill
\begin{minipage}{0.43\linewidth}
\psfig{figure=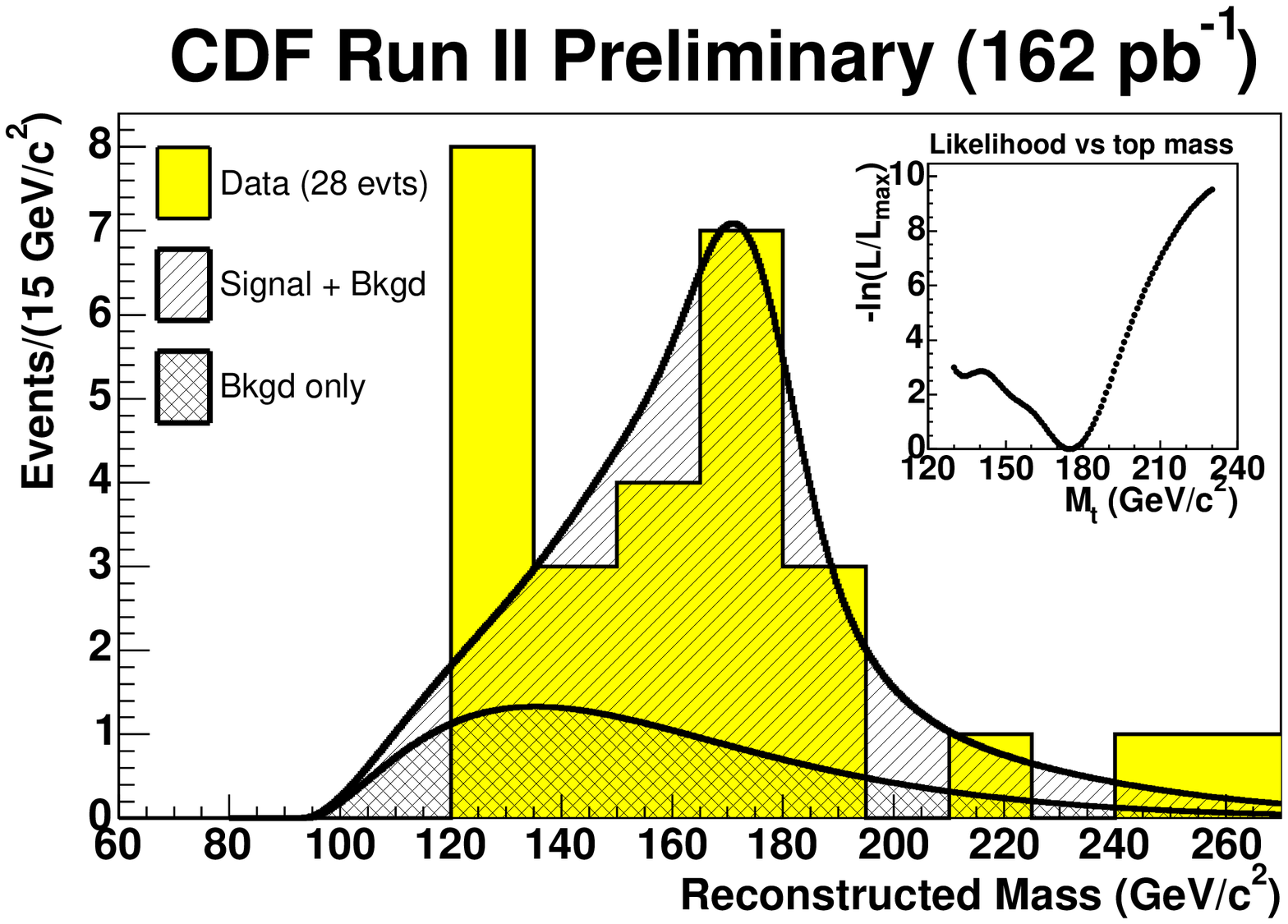,height=2.5in, width=\linewidth}
\end{minipage} \hfill
\caption{Reconstructed top mass distribution for the observed events in 
dilepton (left) and tagged lepton+jets (right) samples. Templates for signal 
and background are also shown. The inner figures show the likelihood as a
function of $m_t$. \label{fig:template}}
\end{figure}

The top mass has also been measured using "template" analyses, in which a 
value for the top mass is reconstructed for each event by looping over all 
possible jet-parton assignments and neutrino solutions, imposing kinematic
constrains, and choosing the $m_t$ which best fits the event. The resulting
$m_t$ distribution is then compared to Monte Carlo $m_t$ templates simulated
at various top masses. The top mass is determined by maximizing a likelihood 
built as a function of $m_t$ by fitting the templates to the observed 
distribution. This method has been used in both dilepton and lepton+jets
samples. Figure~\ref{fig:template} shows the reconstructed top mass 
distributions for the dilepton sample, which observes 12 events and uses 
$\approx$ 193 \pb, and for the tagged lepton+jets sample, which uses
$\approx$ 162 \pb\ and observes 28 events. The figure also shows signal and
background templates and the likelihood as a function of $m_t$.
The corresponding top mass is determined to be
\bea
m_t &=& 176.5^{+17.2}_{-16.0}\rm{(stat)} \pm 6.9\rm{(sys) \ GeV } \nonumber \\
m_t &=& 174.8^{+7.1}_{-7.7}\rm{(stat)} \pm 6.5\rm{(sys) \ GeV } \nonumber
\eea
for the dilepton and lepton+jets channels, respectively.

Several other measurements have been performed, including multivariate
templates, neutrino weighting, and double tags among others.

\section{Single Top}
\begin{figure}
\begin{minipage}{0.43\linewidth}
\psfig{figure=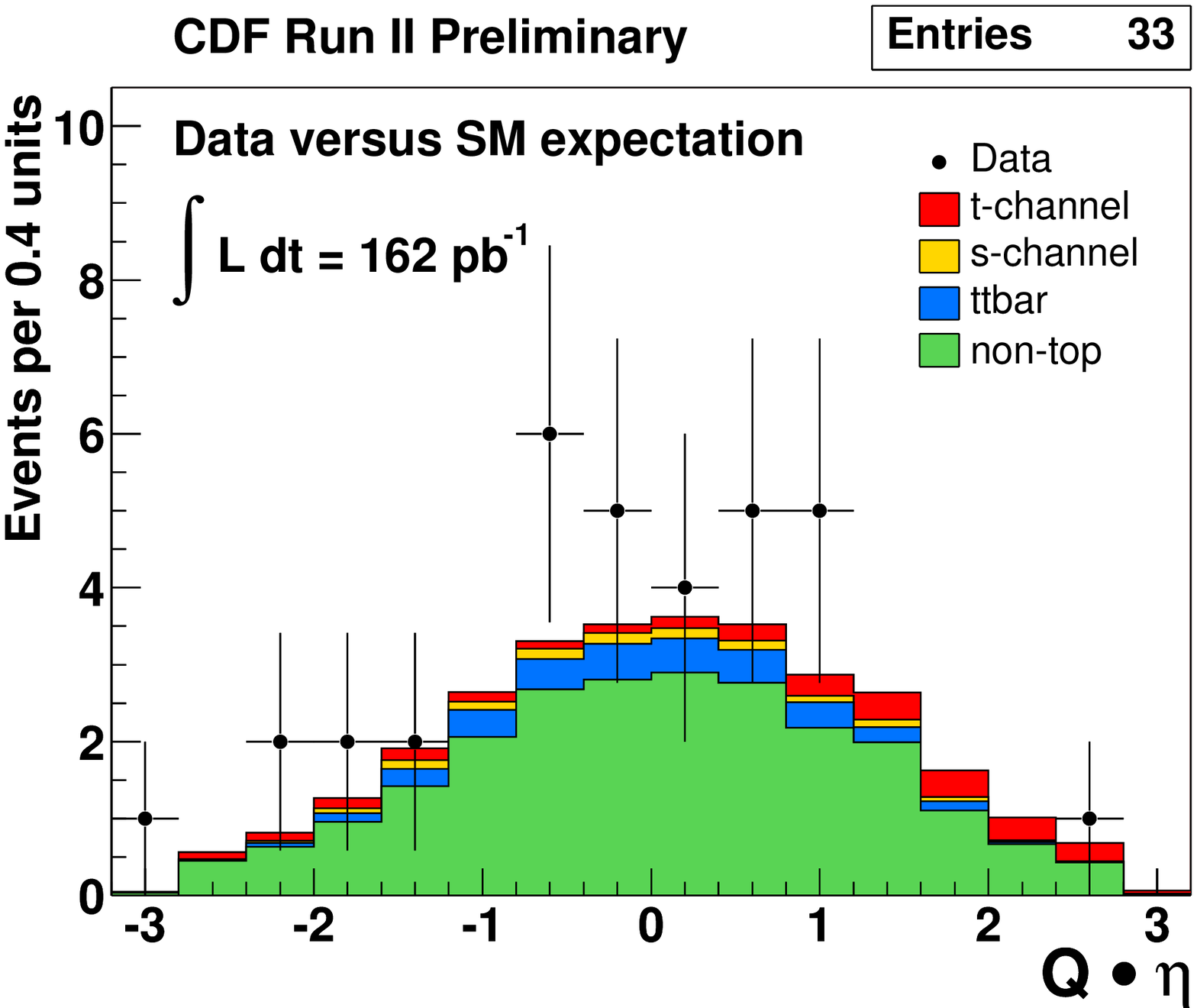, height=2.5in, width=\linewidth}
\end{minipage} \hfill
\begin{minipage}{0.43\linewidth}
\psfig{figure=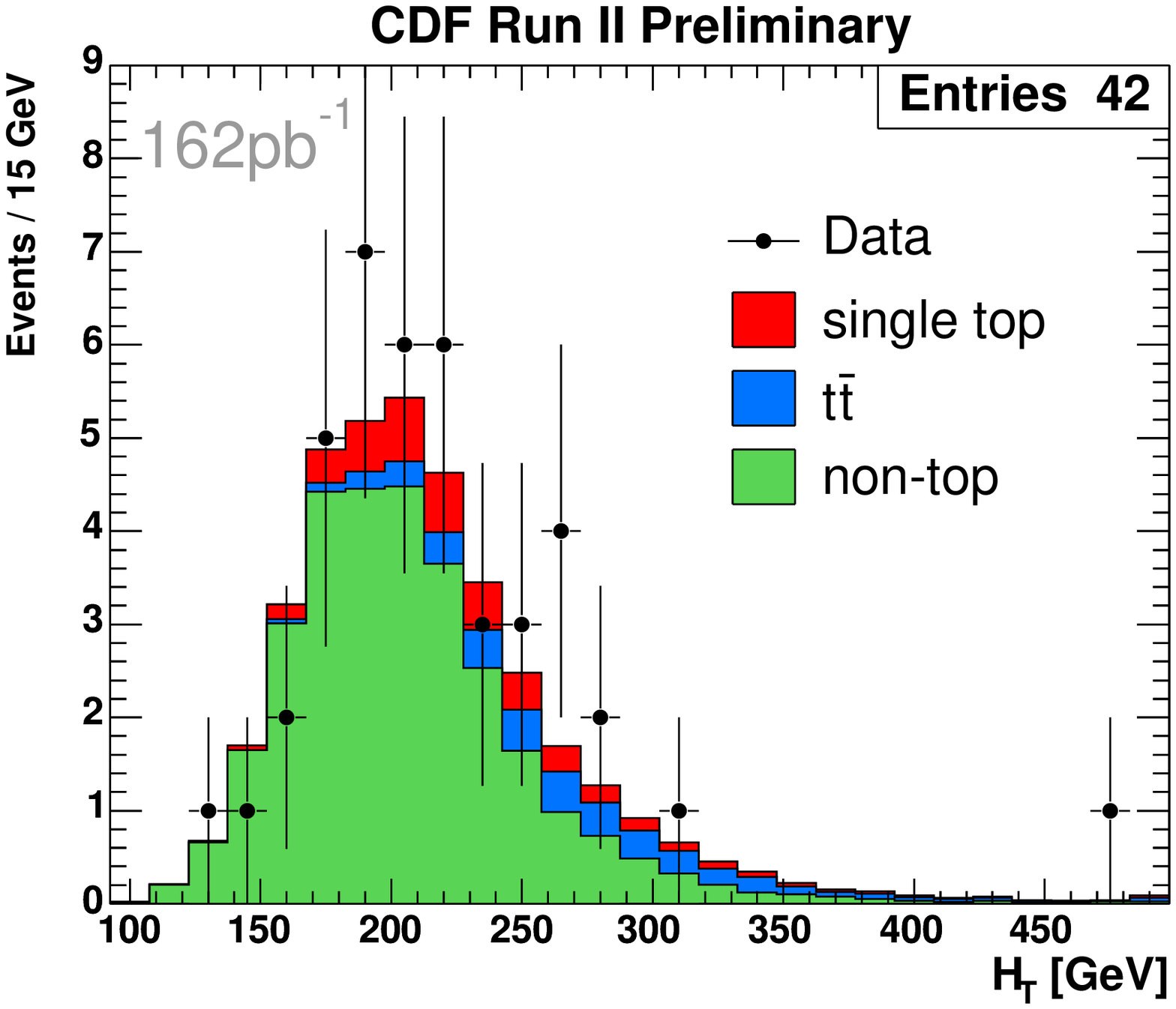,height=2.5in, width=\linewidth}
\end{minipage} \hfill
\caption{The $Q\cdot\eta$ distribution (left) and the $H_T$ distribution 
(right) for the observed single top candidates. The solid histograms are the 
signal and background templates fit to the observed distribution. 
\label{fig:st}}
\end{figure}

Top quarks at the Tevatron can be singly produced via the weak interaction
involving a $Wtb$ vertex. The two relevant production modes are the
$t$ and $s$ channel exchange of a virtual $W$ boson. The predicted theoretical 
cross sections~\cite{single-top} are $\sigma_t = 0.88 \pm 0.11$ pb for the $s$ 
channel and $\sigma_t = 1.98 \pm 0.25$ pb for the $t$ channel.
The single top cross section is of particular interest because it is 
proportional to $|V_{tb}|^2$, the square of the CKM matrix element which 
relates top and bottom quarks, and which has not been directly measured.
Anomalously high rates of single top production would be a hint for exotic 
physics beyond the SM such a FCNC, $W^{\prime}$, or anomalous couplings.
In addition, single top is an important background for SM Higgs searches, since
the final state is the same as for $WH \rightarrow l\nu b\bar{b}$.
The event selection requires an identified $e$ or $mu$, high missing $E_T$, 
and two jets, one of which must be $b$-tagged. Two separate searches are 
performed using $\approx$ 162 \pb\ of data: a combined search for the
$s$ and $t$ channel single top signal, and a separate search which measures the
rates of the two single top processes separately. A maximum likelihood 
technique is used to extract the signal content in the data. Monte Carlo
templates of an appropriate kinematic variable are built for the signal and 
for the \tt\ and non-top backgrounds, and fit to the distribution observed in 
the data. The separate search uses a $Q\cdot\eta$ distribution which exhibits
a distinct asymmetry for $t$ channel events, where $Q$ is the charge of the
lepton and $\eta$ is the pseudorapidity of the untagged jet. 
The combined search uses the $H_T$ distribution. 
The results of the fits to the data are shown in Figure~\ref{fig:st}.
No significant evidence for single top is observed in the data, and 95\% C.L.
upper limits are set. For the separate search, we set
\bea
\sigma_s^{95\% CL} < 13.6 \ \rm{pb} \nonumber \\
\sigma_t^{95\% CL} < 10.1 \ \rm{pb} \nonumber
\eea
and for the combined search we set
\bea
\sigma_{s+t}^{95\% CL} < 17.8 \ \rm{pb} \nonumber
\eea
CDF hopes to see evidence of single top production and to measure its 
production rate with a much larger data sample.

\section{Other Top Measurements}
Several other interesting top quark measurements have been and are being 
performed. Because $m_t > m_W$, a large fraction $F_0$ of $W$ bosons produced 
in top decays are longitudinally polarized. The SM tree-level prediction is
$F_0 = 0.703$ for $m_t = 175$ GeV. This fraction can be measured because
kinematic distributions of the decay products of the $W$ are different for the
different $W$ polarization states. Fitting the $\cos\theta^*$ distribution,
where $\theta^*$ is the angle between the charged lepton momentum in the $W$
rest frame and the boost direction from the top to the $W$ rest frame,
in a 162 \pb\ tagged lepton+jets sample we determine 
$F_0 = 0.89^{+0.30}_{-0.34}$(stat)$\pm 0.17$(sys) and 
$F_0 > 0.25$ \@ 95\% C.L. 
A similar analysis uses the $p_T$ of the lepton to discriminate between 
different $W$ polarizations. In the tagged lepton+jets sample the result is
$F_0 = 0.88^{+0.12}_{-0.47}$(tot) and 
$F_0 > 0.24$ \@ 95\% C.L. 
In a 200 \pb\ dilepton sample we determine 
$F_0 < 0.52$ \@ 95\% C.L. 
and the combined analysis gives
$F_0 = 0.27^{+0.35}_{-0.24}$(stat)$\pm 0.17$(sys) and 
$F_0 < 0.88$ \@ 95\% C.L. 
The reason for the dilepton result is an excess of events with low $p_T$
leptons which drives the maximum likelihood for $F_0$ to an unphysical 
negative value.

A search for anomalous kinematics in top decays has been performed. All
distributions agree with SM expectations except for the above mentioned excess 
of low $p_T$ dilepton events, which is compatible with the SM with a 1\% to 4\%
probability.

Top events are an interesting sample to search for physics beyond the SM.
A search for charged Higgs from top decays sets tree-level limits in the MSSM
($m_H$,$\tan\beta$) plane which improve upon previous existing limits, and
a model independent limit of BR($t\rightarrow Hb$)$<$0.7 at 95\% C.L. is set
on the top to Higgs branching fraction for charged Higgs masses of
80 GeV $< m_H <$ 150 GeV. 
Another search for a generic 4$^th$ generation $t^{\prime}$ quark is performed
using 195 \pb\ of data by fitting the $H_T$ distribution in the untagged 
lepton + $\geq 4$ jet  events to
different background, top, and $t^{\prime}$ signal templates generated at
various $t^{\prime}$ masses. Upper limits at 95\% C.L. are set on the
$t^{\prime}$ cross section as a function of $t^{\prime}$ mass. The projected
sensitivity of this analysis predicts setting limits on the $t^{\prime}$ mass
with an integrated luminosity greater than 500 \pb. 

Several branching fraction studies were performed. The rate of tau leptons
in top decays is found to be in agreement with SM predictions, setting 
a limit of $r_{\tau} < 5.0$ at the 95\% C.L. for the anomalous rate enhancement
factor, predicted to be unity by the SM. The ratio 
$ R = $ BR($t\rightarrow Wb$)/BR($t\rightarrow Wq$) is measured in both 
dilepton and lepton+jets events, setting a limit of $R > 0.62$ at 95\% C.L., in
agreement with the SM prediction of $R > 0.998$. The ratio of dilepton to
lepton+jets cross-sections is measured to be 
$\sigma_{ll}/\sigma_{lj} = 1.45^{+0.83}_{-0.55}$, in agreement with the 
expected value of 1.

\section{Conclusions and Outlook}
Top quark properties are under extensive study at CDF. The \tt\ production
cross section has been measured in different top decay channels and using 
different techniques. All results are in agreement with one another and with
the theoretical predictions, giving us confidence that indeed we have 
established a top signal and that we understand its backgrounds, its heavy
flavor composition, and its kinematic properties. The top mass has been
measured using different samples and techniques. The largest systematic 
uncertainty in these measurements comes from the jet energy scale 
determination, and much improved top mass results will be shown shortly after
this conference. With larger data samples and improved jet energy measurements,
the overall top mass uncertainty will soon be reduced to about 4 GeV.
We have searched for electroweak production of top, setting limits on its
production cross section. With much larger data samples we hope to observe
a single top signal and measure its cross section.
By the end of 2007, the recorded data will reach at least 2 fb$^{-1}$, and it
could be more than double this size depending on Tevatron performance.
All analyses will reduce significantly their uncertainties, and searches for
new physics will significantly improve their sensitivity. 

\section*{Acknowledgments}
The results shown here represent a lot of work by a lot of people.
I thank my CDF colleagues for their efforts to carry out these challenging
physics analyses and for providing a friendly and stimulating research 
environment. I thank the conference organizers for a very nice and well 
organized week of physics. Finally, I thank the colleagues of my research
institution, IFCA, for their support. 
This work is supported by the Ministerio de Educaci\'on y Ciencia, Spain 
(under contract FPA2002-01678) and by the EU 
(under contract HPRN-CT-2002-00292).


\section*{References}
\begin{thebibliography}{99}

\bibitem{cdf} The CDF Collaboration, {\em The CDF II Detector Technical Design
Report}, FERMILAB-Pub-96/390-E.

\bibitem{xsec-th} M. Cacciari, S. Frixione, G. Ridolfi, M. Mangano and
P. Nason, {\it JHEP} {\bf404}, 68 (2004).

\bibitem{single-top} B.W. Harris, E. Laenen, L. Phaf, Z. Sullivan, 
S. Weinzierl, \Journal{\PRD}{66}{054024}{2002}; Z. Sullivan, hep-ph/0408049.

\end{thebibliography}
\end{document}